# Analysis of the Sufficient Path Elimination Window for the Maximum-Likelihood Sequential-Search Decoding Algorithm for Binary Convolutional Codes

Shin-Lin Shieh[†‡], Po-Ning Chen[‡] and Yunghsiang S. Han[♯]

**Abstract**

In this work, the priority-first sequential-search decoding algorithm proposed in [8] is revisited. By replacing the conventional Fano metric by one that is derived based on the Wagner rule, the sequential-search decoding in [8] guarantees the maximum-likelihood (ML) performance, and hence, was named the *maximum-likelihood sequential decoding algorithm* (MLSDA). It was then concluded by simulations that the software computational complexity of the MLSDA is in general considerably smaller than that of the Viterbi algorithm.

A common problem on sequential-type decoding is that at the signal-to-noise ratio (SNR) below the one corresponding to the cutoff rate, the average decoding complexity per information bit and the required stack size grow rapidly with the information length [13]. This problem to some extent prevent the practical use of sequential-type decoding from convolutional codes with long information sequence at low SNRs. In order to alleviate the problem in the MLSDA, we propose to directly eliminate the top path whose end node is $\Delta$-trellis-level prior to the farthest one among all nodes that have been expanded thus far by the sequential search, which we termed the *early elimination*. Following random coding argument, we analyze the early-elimination window $\Delta$ that results in negligible performance degradation

[†]SunplusMM Technology Co., Ltd, HsinChu City, Taiwan 300, ROC

[‡]Dept. of Communications Eng., National Chiao-Tung Univ., HsinChu City, Taiwan 300, ROC

[♯]Graduate Institute of Communication Eng., National Taipei Univ., Taipei, Taiwan 237, ROC

for the MLSDA. Our analytical results indicate that the required early elimination window for negligible performance degradation is just twice of the constraint length for rate one-half convolutional codes. For rate one-third convolutional codes, the required early-elimination window even reduces to the constraint length. The suggestive theoretical level thresholds almost coincide with the simulation results. As a consequence of the small early-elimination window required for near maximum-likelihood performance, the MLSDA with early-elimination modification rules out considerable computational burdens, as well as memory requirement, by directly eliminating a big number of the top paths, which makes the MLSDA with early elimination very suitable for applications that dictate a low-complexity software implementation with near maximum-likelihood performance.

**Index Terms**

Sequential decoding, maximum-likelihood, soft-decision, random coding

## I. Introduction

One of the most commonly used decoding algorithms for convolutional codes is the Viterbi algorithm. It operates on a convolutional code trellis, and has been shown to be a maximum-likelihood decoder [13]. Since its decoding complexity grows exponentially with the code constraint length, the Viterbi algorithm is usually applied only for convolutional codes with short constraint lengths.

When the information sequence is long, path truncation was suggested for a practical implementation of the Viterbi decoder [13]. Instead of keeping all trellis branches on the survivor paths in the decoder memory, only a certain number of the most recently trellis branches is retained, and a decision is forced on the oldest trellis branch whenever a new data arrives the decoder. In literatures, three strategies have been proposed on the forceful decision: (1) majority vote strategy that traces back from all states and outputs the decision that occurs most often; (2) best state strategy that only traces back from the state with the best metric and outputs the information bits corresponding to the path being traced; (3) random state strategy that randomly traces back from one state and outputs the information bits corresponding to the path being traced. Although none of the three forceful strategies guarantees maximum-likelihood, their performance degradation can be made negligible as long as the traceback window or truncation window is sufficiently large.

In [3], Forney proved that a truncation window of $5.8$-fold of the code constraint length suffices to provide negligible performance degradation for the best state strategy. Hemmati and Costello [11] later derived an upper performance bound as a function of the truncation window, and obtained a similar conclusion for the best state strategy. McEliece and Onyszchuk [14] studied the tradeoff between length of the truncation window and performance loss for the random state strategy, and concluded that the truncation window for the random state strategy should be about twice as large as that for the best state strategy.

On the other hand, the sequential decoding algorithm has received little attention in the past 30 years due to its sub-optimum performance and lack of efficient and cost-effective hardware implementation. However, because its decoding complexity is irrelevant to the code constraint length, the sequential decoding algorithm is suitable for convolutional codes with large memory order. For this reason, it has been recently proposed to be used in the decoding of the so-called "super-code" that considers the joint effect of multi-path channel and convolutional codes [10].

Based on the Wagner rule, a variant of the sequential decoding algorithm has been established, and was proved to be maximum-likelihood [8]. As a result, the new sequential-type decoding algorithm is termed the *maximum-likelihood sequential decoding algorithm* (MLSDA) for referring convenience. By simulations, the authors in [8] observed that, from pure software implementation standpoint, the average decoding complexity of the MLSDA is in general considerably smaller than the Viterbi algorithm when the signal-to-noise ratio (SNR) of the additive white Gaussian noise (AWGN) channel is larger than 2 dB.

Similar to the Viterbi algorithm, the decoding burdens of the sequential decoding algorithm, both in memory consumption and in computational complexity, grows as the length of the infor-mation sequence increases. However, in order to compensate the SNR loss due to the additional zeros at the end of the information sequence, a long information sequence is often preferred in practice. One solution to reduce the decoding burdens as a result of a long information sequence is to introduce the path truncation concept of the Viterbi algorithm to the sequential decoding algorithm. As one example, Zigangirov [19] derived an error probability upper bound of the sequential decoding with backsearch limit, in which the decoder traces back the top path in stack to output the forceful decisions of the symbols at those levels prior to the backsearch limit. Under the situation that the channel critical rate is smaller than $(\kappa-1)/\kappa$ of the computational cutoff rate, where $\kappa$ is the ratio of backsearch limit against code constraint length, Zigangirov's bound was shown to reduce to the Yudkin-Viterbi bound [5] for infinite backsearch limit at low to medium rates, and to coincide with the random coding bound at high rate [19].

In this paper, an alternative approach to lower the decoding complexity of the sequential decoding is examined. Instead of tracing back the top path in stack to force the decision of the symbols beyond the backsearch limit, we propose to directly eliminate the top path whose end node is $\Delta$-level-prior to the farthest node among all nodes that have been expanded thus far by the sequential search, which is named the *early elimination*. Following similar random coding argument used by Forney [3], we analyze the early-elimination window $\Delta$ that results in negligible performance degradation for the MLSDA. Our analytical results indicate that the required early elimination window for negligible performance degradation is just twice of the constraint length for rate one-half convolutional codes. For rate one-third convolutional codes, the required early-elimination window even reduces to the constraint length. Simulations are also performed, and they confirm and match the analytical results.

As a consequence of the small early-elimination window required for near maximum-likelihood performance, the MLSDA with early-elimination modification rules out considerable computational burdens, as well as memory requirement, by directly eliminating a big number of the top paths. Furthermore, it can also be implemented together with the backsearch scheme to provide timely decision of fixed delay, and to further reduce the decoding complexity. This suggests the potential and suitability of the MLSDA with early elimination for applications that dictate a low-complexity software implementation with near maximum-likelihood performance.

The rest of the paper is organized as follows. The preliminary results are briefed in Section II. The early elimination modification of the MLSDA is presented in Section III. The analysis of the sufficient early elimination window for near-maximum-likelihood performance is given in Section IV. Numerical and Simulation results are remarked in Section V. Section VI concludes this paper.

Throughout the paper, natural logarithm is assumed except otherwise stated.

## II. Preliminaries

In this section, we present the system model considered in this work. The technique that Forney used to prove that a truncation window of $5.8$-fold of the code constraint length is sufficient to secure near-optimal performance for the best state strategy is introduced in brief for completeness.

### A. System Model and the MLSDA

Let $\mathcal{C}$ be a binary $(n, k, m)$ convolutional code with finite input information sequence of $k \times L$ bits, followed by $k \times m$ zeros to clear the encoder memory. Thus, $\mathcal{C}$ forms an $(N, K)$ linear block code with effective code rate $R = K/N$, where $K = kL$ and $N = n(L+m)$. Denote the parity check matrix of $\mathcal{C}$ by $\mathbb{H}$. The code rate, the memory order and the constraint length of $\mathcal{C}$ are given by $k/n$, $m$ and $(m+1)$, respectively. Put a binary codeword of $\mathcal{C}$ by $\boldsymbol{v} \triangleq (v_0, v_1, \ldots, v_{N-1})$, where each $v_j \in \{0, 1\}$. For notational convenience, we represent a portion of codeword $\boldsymbol{v}$ by $\boldsymbol{v}_{(a,b)} \triangleq (v_a, v_{a+1}, \ldots, v_b)$, and abbreviate $\boldsymbol{v}_{(0,b)}$ and $\boldsymbol{v}_{(0,N-1)}$ as $\boldsymbol{v}_{(b)}$ and $\boldsymbol{v}$, respectively. Same abbreviation will be applied to other vector notations.

Assume that the binary codeword is transmitted over a binary-input time-discrete channel with channel output $\boldsymbol{r} \triangleq (r_0, r_1, \ldots, r_{N-1})$. Define hard-decision sequence $\boldsymbol{y} \triangleq (y_0, y_1, \ldots, y_{N-1})$ corresponding to $\boldsymbol{r}$ as:

$$y_j \triangleq \begin{cases} 1, & \text{if } \phi_j < 0; \\ 0, & \text{otherwise,} \end{cases}$$

where $\phi_j \triangleq \log[\Pr(r_j|v_j = 0)/\Pr(r_j|v_j = 1)]$, and $\Pr(r_j|v_j)$ denotes the channel transition probability of $r_j$ given $v_j$. According to the Wagner rule, the maximum-likelihood decoding output $\hat{\boldsymbol{v}}$ for received vector $\boldsymbol{r}$ can be obtained by

$$\hat{\boldsymbol{v}} = \boldsymbol{y} \oplus \boldsymbol{e}^*, \tag{1}$$

where "$\oplus$" is the bit-wise exclusive-or operation, and $\boldsymbol{e}^*$ is the one with the smallest $\sum_{j=0}^{N-1} e_j|\phi_j|$ among all error patterns $\boldsymbol{e} \in \{0,1\}^N$ satisfying $\boldsymbol{e}\mathbb{H}^T = \boldsymbol{y}\mathbb{H}^T$. Here, superscript "$T$" is used to denote the matrix transpose operation. Based on the observation in (1), a new sequential-type decoder can be established by replacing the Fano metric in the conventional sequential decoding algorithm by a metric defined as:

$$\mu\left(\boldsymbol{x}_{(\ell n-1)}\right) \triangleq \sum_{j=0}^{\ell n-1} \mu(x_j). \tag{2}$$

where $\boldsymbol{x}_{(\ell n-1)} = (x_0, x_1, \ldots, x_{\ell n-1}) \in \{0,1\}^{\ell n}$ represents the label of a path ending at level $\ell$ in the $(n, k, m)$ convolutional code tree, and $\mu(x_j) \triangleq (y_j \oplus x_j)|\phi_j|$. Since the new decoding metric is nondecreasing along the code path, and since finding $\boldsymbol{e}^*$ is equivalent to finding the code path with the smallest metric in the code tree, it was proved in [8] that the new sequential-type decoder can always locate the maximum-likelihood codeword through the greedy-in-nature priority-first sequential codeword search. For this reason, the new sequential-type decoder is named the *maximum-likelihood sequential decoding algorithm* (MLSDA) [8].

By adding another stack, the MLSDA can be made to operate on a code trellis instead of a code tree [8]. The two stacks used in the trellis-based MLSDA are referred to as the *Open Stack* and the *Closed Stack*. The *Open Stack* contains all paths that end at the frontier part of the trellis being thus far explored (cf. Fig. 2). The Open Stack functions similarly as the single stack in the conventional sequential decoding algorithm. The *Closed Stack* stores the information of the ending states and ending levels of the paths that had been the top paths of the Open Stack. The Closed Stack is used to determine whether two paths intersect in the code trellis during the sequential search. The trellis-based MLSDA [8] is quoted below for completeness.

**<Trellis-Based MLSDA>**

*Step 1. Load the Open Stack with the origin node whose metric is zero.*

*Step 2. Put into the* Closed Stack *both the state and level of the end node of the top path in the* Open Stack. *Compute the path metric for each of the successor paths of the top path in the* Open Stack *by adding the branch metric of the extended branch to the path metric of the top path. Delete the top path from the* Open Stack.

*Step 3. Discard the successor paths in Step 2, which end at a node that has the same state and level as any entry in the* Closed Stack. *If any successor path ends at the same node as a path already in the* Open Stack*, eliminate the path with higher path metric.*[1]

*Step 4. Insert the remaining successor paths into the* Open Stack *in order of ascending path metrics. If two paths in the Open Stack have equal metric, sort them in order of descending levels. If, in addition, they happen to end at the same level, sort them randomly.*

*Step 5. If the top path in the* Open Stack *reaches the end of the convolutional code trellis, the algorithm stops; otherwise go to Step 2.*

It is known that the computational efforts of the sequential-search decoding algorithms, including the trellis-based MLSDA, are determined not only by the number of metrics computed, but also by the cost of searching and inserting of the stack elements. However, the latter cost can be made of comparable order to the former by adopting the *double-ended heap* (DEAP) [2] data structure in the stack implementation. This justifies the common usage of number of metric computations as the key determinant of the algorithmic complexity of the sequential-search decoding algorithm.

[1]For discrete channels, it may occur that the successor path not only ends at the same node as some path already in the Open Stack but also has equal path metric to it. In such case, randomly eliminate one of them.

### *B. Random Coding Analysis of the Path Truncation Window*

In [4], Gallager considered the discrete memoryless channel with input alphabet size $I$, output alphabet size $J$ and channel transition probability $P_{ji}$, and presented the random coding bound for the maximum-likelihood decoding error $P_e$ of the $(N, K)$ block code as:

$$P_e \leq \exp\left\{-N\left[-\rho R + E_0(\rho, \boldsymbol{p})\right]\right\}$$

for all $0 \leq \rho \leq 1$, where $R = \log(I^K)/N = (K/N)\log(I)$ is the code rate measured in nats per symbol, $\boldsymbol{p} = (p_1, p_2, \cdots, p_I)$ is the input distribution adopted for the random selection of codewords, and

$$E_0(\rho, \boldsymbol{p}) \triangleq -\log \sum_{j=1}^{J} \left( \sum_{i=1}^{I} p_i P_{ji}^{1/(1+\rho)} \right)^{1+\rho}. \tag{3}$$

Gallager's result leads to the well-known random coding exponent:

$$E_r(R) \triangleq \max_{0\leq\rho\leq 1} \max_{\boldsymbol{p}} [-\rho R + E_0(\rho, \boldsymbol{p})] = \max_{0\leq\rho\leq 1} [-\rho R + E_0(\rho)],$$

where $E_0(\rho) \triangleq \max_{\boldsymbol{p}} E_0(\rho, \boldsymbol{p})$ is the Gallager function [20]. Notably, the random coding exponent is a lower bound of the channel reliability function $E(R) \triangleq \lim_{N\to\infty} -(1/N)\log(P_e)$ (provided the limit exists), and is tight for code rates above the cutoff rate.

In [17], Viterbi applied similar random coding argument to the derivation of the decoding error for time-varying convolutional codes. Specifically, he considered a single-input $n$-output convolutional encoder with one $(m + 1)$-stage shift register as shown in Fig. 1. The $n$ inner product computers may change with each new input symbol, and hence, a time-varying code trellis is resulted. As all elements are assumed to be in $GF(q)$, each input symbol will induce $q$ branches on the code trellis, and each branch is labelled by $n$ channel symbols. As a result of the attached $m$ zeros at the end, the encoder will produce $n(L+m)$ output channel symbols in response to the input sequence of $L$ symbols. Under the above system setting, Viterbi showed that the maximum-likelihood decoding error $P_{e,c}$ for time-varying convolutional codes can be upper-bounded by:

$$P_{e,c} \le \frac{q-1}{1-q^{-\lambda/R}} \exp[-n(m+1)E_0(\rho)] \tag{4}$$

for all $0 \le \rho \le 1$, where $R \triangleq \log(q)/n$ is the code rate in unit of nats per symbol, and $\lambda \triangleq E_0(\rho) - \rho R$ is a constant independent of $n(m+1)$. Since $\lambda$ is required to be positive, it can be concluded that:

$$\liminf_{n\to\infty} -\frac{1}{n} \log P_{e,c} \ge (m+1)E_c(R),$$

where $E_c(R) \triangleq \max_{\{\rho\in[0,1]\ :\ E_0(\rho)>\rho R\}} E_0(\rho)$. For symmetric channels, $E_0(\rho)$ is an increasing and concave function in $\rho$ with $E_0(0) = 0$; therefore, $E_c(R)$ can be reduced to:

$$E_c(R) = \begin{cases} R_0, & \text{if } 0 \le R < R_0; \\ E_0(\rho^*), & \text{if } R_0 \le R < C; \\ 0, & \text{if } R \ge C, \end{cases} \tag{5}$$

where $R_0 = E_0(1)$ is the cutoff rate, $C = E_0'(0)$ is the channel capacity, and $\rho^* = \rho^*(R)$ is the unique solution of $E_0(\rho) = \rho R$. It is also shown in the same work that $E_c(R)$ is a tight exponent for $R \ge R_0$.

In order to derive the path truncation window with near-optimal performance, Forney [3] treated the truncated convolutional code as a block code, and upper-bounded the additional decoding error $P_{e,T}$ due to path truncation in the Viterbi decoder by means of Gallager's technique as:

$$P_{e,T} \le \exp[-n\tau E_r(R)], \tag{6}$$

where $\tau$ is the truncation window size. Forney then noticed that as long as

$$\liminf_{n\to\infty} -\frac{1}{n} \log P_{e,T} > \limsup_{n\to\infty} -\frac{1}{n} \log P_{e,c}, \tag{7}$$

the additional error $P_{e,T}$ due to path truncation becomes exponentially negligible with respect to $P_{e,c}$. For $R \geq R_0$, condition (7) reduces to

$$\tau E_r(R) > (m+1)E_c(R)$$

by inequality (6) and the tightness of $E_c(R)$. A specific case is given in Fig. 3 in which the binary symmetric channel (BSC) with crossover probability $0.4$ gives that the path truncation window at the cutoff rate $R_0 = 0.0146$ bit/symbol must be larger than $E_c(R_0)/E_r(R_0) \approx 0.0146/0.0025 = 5.84$-fold of the code constraint length. This number parallels the one obtained under the very noisy channels, where 5.8-fold of the code constraint length is suggested for the path truncation window at the cutoff rate [18].

## III. Early Elimination Scheme for Priority-First Sequential-Search algorithm

The early elimination modification that we proposed in this work is based on the following observation. As shown in Fig. 2, suppose that the path ending at node $C$ is a portion of the final code path to be located at the end of the sequential search, and suppose that the path ending at node $D$ happens to be the current top path. Then, expanding node $D$ until all of its offsprings finally have decoding metrics exceeding those of the successors of the path ending at node $C$ may consume considerable but unnecessary number of computational efforts. This observation hints that by setting a proper level threshold $\Delta$ and directly eliminating the top path whose level is no larger than $(\ell_{\max} - \Delta)$, where $\ell_{\max}$ is the largest level for all nodes that have been expanded thus far by the sequential search, the computational complexity of the priority-first sequential-search algorithm may be reduced without sacrificing much of the performance.

It is worth mentioning that if the decoding metric is monotonically nondecreasing along the path portion to be searched, it can be guaranteed that the path that updates the current $\ell_{\max}$ is always the one with the smallest path metric among all paths ending at the same level [8]. This is the key to lead to the result that the sequential search using a monotonic maximum-likelihood metric like (2) can assure that the first top path that reaches the last level of the code tree or code trellis is exactly the maximum-likelihood code path.

Based on the above observation, we propose to set a level threshold $\Delta$ in the trellis-based MLSDA, and directly eliminate the top path whose level is no larger than $(\ell_{\max} - \Delta)$. For this modification, we only need to modify Step 2 in the trellis-based MLSDA as follows.

**<Trellis-Based MLSDA with Early Elimination Modification>**

*Initialization. Set a level threshold* $\Delta$*. Assign* $\ell_{\max} = 0$.

*Step 2*$'$*. Perform the following check before executing the original Step 2 in the trellis-based MLSDA.*

- *If the top path in the Open Stack ends at a node whose level is no larger than* $(\ell_{\max} - \Delta)$*, directly eliminate the top path, and go to Step 5; otherwise, update* $\ell_{\max}$ *if it is smaller than the ending level of the current top path.*

## IV. Analysis of the Early-Elimination Window with Negligible Performance Degradation

This section provides detailed derivation on the early elimination window that yields negligible performance degradation.

Referring to Fig. 2, suppose that the path that ends at node $B$ is the current top path of the Open Stack. Let the current $\ell_{\max}$ be updated due to the expansion of node $C$. According to the merging operation performed at Step 3 of the trellis-based MLSDA, all the paths survived in the Open Stack should follow distinct traces except possibly for forward few branches.[2] Hence, we

[2]One example of the claimed statement is that the path ending at node $B$ and the path ending at node $D$ have distinct traces after node $G$, and share common branches all the way before node $G$. This property applies to all the pathes that end at the unfilled-circle node in Fig. 2.

Note that edges $EQ$ and $HF$ were drawn in dotted lines in Fig. 2 because after merging, no pathes in the Open Stack pass through them.

may assume that the path that ends at node $B$ and the path that updates the current $\ell_{\max}$ have common traces before node $A$, whose level is denoted by $\ell_{\min}$ for convenience. Let $\boldsymbol{x}_{(\ell_{\min}n-1,\ell n-1)}$ and $\tilde{\boldsymbol{x}}_{(\ell_{\min}n-1,\ell_{\max}n-1)}$ be the labels corresponding to path portions $AB$ and $AC$, respectively. With the above setting, we are interested in the additional decoding probability error per node (i.e., node $A$) due to early elimination as following [20]. Without loss of generality, set $\ell_{\min}=0$ in the below derivation.

Observe that the current top path ending at node $B$ is early-eliminated if, and only if, node $C$ is expanded earlier than node $B$, provided $\ell \le (\ell_{\max}-\Delta)$. Since the decoding metric of the MLSDA is monotonically nondecreasing along the path portion to be searched, that node $C$ is expanded prior to node $B$ is equivalent to the condition that $\mu\left(\boldsymbol{x}_{(\ell n-1)}\right) \ge \mu\left(\tilde{\boldsymbol{x}}_{(\ell_{\max}n-1)}\right)$, which, according to the maximum-likelihoodness of the metrics, is in turn equivalent to:

$$\Pr\left(\boldsymbol{r}_{(\ell n-1)}\middle|\,\boldsymbol{x}_{(\ell n-1)}\right) \le \Pr\left(\boldsymbol{r}_{(\ell_{\max}n-1)}\middle|\,\tilde{\boldsymbol{x}}_{(\ell_{\max}n-1)}\right). \tag{8}$$

By noting that for the MLSDA, the path that updates the current $\ell_{\max}$ is exactly the one with the smallest path metric among all paths ending at the same level [8], condition (8) can be equivalently re-written as:

$$\Pr\left(\boldsymbol{r}_{(\ell n-1)}\middle|\,\boldsymbol{x}_{(\ell n-1)}\right) \le \max_{\tilde{\boldsymbol{x}}_{(\ell_{\max}n-1)}\in\mathscr{C}_{\ell_{\max}}} \Pr\left(\boldsymbol{r}_{(\ell_{\max}n-1)}\middle|\,\tilde{\boldsymbol{x}}_{(\ell_{\max}n-1)}\right), \tag{9}$$

where $\mathscr{C}_{\ell_{\max}}$ is the set of all labels of length $\ell_{\max}n$, whose corresponding paths consist of different branches from path $AB$ after node $A$. Consequently, additional decoding error may be introduced by early elimination if (9) is valid for some $\ell$ and $\ell_{\max}$ with $\ell \le (\ell_{\max}-\Delta)$, when $\boldsymbol{x}$ is the transmitted codeword.[3]

[3]Since the early-elimination of path with label $\boldsymbol{x}$ is always performed whenever (9) is valid, it is clear that additional error is introduced only when the transmitted label $\boldsymbol{x}$ corresponds to the maximum-likelihood code path. In other words, when $\boldsymbol{x}$ does not label the maximum-likelihood code path, the validity of (9) or early-elimination of path with label $\boldsymbol{x}$ will not add a new error to maximum-likelihood decoding. As what we concern is an upper probability bound for the additional error due to early-elimination, it suffices to analyze the probability bound on the occurrence of (9).

Notably, when equality holds in (9), $\boldsymbol{x}$ will still be early-eliminated according to Step 4 of the algorithm.

The analysis of the probability bound on the occurrence of (9) is different from what was done by Forney since it compares the channel posterior probabilities given codeword portions of “unequal” lengthes, while Forney dealt with the channel posterior probabilities for codewords of “equal” length. Refinement on Forney’s derivation is therefore necessary.

For notational convenience, we replace $\ell_{\max}$ by $\beta$ in the following formulas. The probability $\xi(\ell,\beta)$ that (9) occurs is given by:

$$\xi(\ell,\beta) = \sum_{\boldsymbol{r}_{(\beta n-1)}\in\mathcal{R}^{\beta n-1}} \Phi_0\left(\boldsymbol{r}_{(\beta n-1)}\right)\Pr\left(\boldsymbol{r}_{(\beta n-1)}\middle|\,\boldsymbol{x}_{(\beta n-1)}\right), \tag{10}$$

where $\mathcal{R}$ is the (possibly discrete or continuous) generic alphabet of $\boldsymbol{r}$, and

$$\Phi_0\left(\boldsymbol{r}_{(\beta n-1)}\right) \triangleq \begin{cases} 1, & \text{if (9) is valid;} \\ 0, & \text{otherwise .}\end{cases}$$

From

$$\Phi_0\left(\boldsymbol{r}_{(\beta n-1)}\right) \le \left[\frac{\displaystyle\sum_{\tilde{\boldsymbol{x}}_{(\beta n-1)}\in\mathscr{C}_\beta} \Pr\left(\boldsymbol{r}_{(\beta n-1)}\middle|\,\tilde{\boldsymbol{x}}_{(\beta n-1)}\right)^{1/(1+\rho)}}{\Pr\left(\boldsymbol{r}_{(\ell n-1)}\middle|\,\boldsymbol{x}_{(\ell n-1)}\right)^{1/(1+\rho)}}\right]^\rho \quad \text{for } \rho\ge 0,$$

we obtain:

$$\begin{aligned}\xi(\ell,\beta) &\le \sum_{\boldsymbol{r}_{(\beta n-1)}\in\Re^{\beta n-1}} \left[\frac{\displaystyle\sum_{\tilde{\boldsymbol{x}}_{(\beta n-1)}\in\mathscr{C}_\beta} \Pr\left(\boldsymbol{r}_{(\beta n-1)}\middle|\,\tilde{\boldsymbol{x}}_{(\beta n-1)}\right)^{1/(1+\rho)}}{\Pr\left(\boldsymbol{r}_{(\ell n-1)}\middle|\,\boldsymbol{x}_{(\ell n-1)}\right)^{1/(1+\rho)}}\right]^\rho \Pr\left(\boldsymbol{r}_{(\beta n-1)}\middle|\,\boldsymbol{x}_{(\beta n-1)}\right), \\ &= \sum_{\boldsymbol{r}_{(\beta n-1)}\in\Re^{\beta n-1}} \left[\sum_{\tilde{\boldsymbol{x}}_{(\beta n-1)}\in\mathscr{C}_\beta} \Pr\left(\boldsymbol{r}_{(\beta n-1)}\middle|\,\tilde{\boldsymbol{x}}_{(\beta n-1)}\right)^{1/(1+\rho)}\right]^\rho \\ &\qquad \times \Pr\left(\boldsymbol{r}_{(\ell n-1)}\middle|\,\boldsymbol{x}_{(\ell n-1)}\right)^{1/(1+\rho)} \Pr\left(\boldsymbol{r}_{(\ell n,\beta n-1)}\middle|\,\boldsymbol{x}_{(\ell n,\beta n-1)}\right),\end{aligned}$$

where the last step uses $\Pr\left(\boldsymbol{r}_{(\beta n-1)}\middle|\,\boldsymbol{x}_{(\beta n-1)}\right) = \Pr\left(\boldsymbol{r}_{(\ell n,\beta n-1)}\middle|\,\boldsymbol{x}_{(\ell n,\beta n-1)}\right)\Pr\left(\boldsymbol{r}_{(\ell n-1)}\middle|\,\boldsymbol{x}_{(\ell n-1)}\right)$, a consequence of the memoryless property of the channel. Taking expectation of $\xi(\ell,\beta)$ with

respect to random selection of labels (i.e., codewords) of length $(\beta n-1)$ yields that:

$$\begin{aligned}
&\overline{\xi(\ell,\beta)}\\
&\leq \overline{\sum_{\boldsymbol{r}_{(\beta n-1)}\in\Re^{\beta n-1}}\left[\sum_{\tilde{\boldsymbol{x}}_{(\beta n-1)}\in\mathscr{C}_\beta}\Pr\left(\boldsymbol{r}_{(\beta n-1)}\middle|\,\tilde{\boldsymbol{x}}_{(\beta n-1)}\right)^{1/(1+\rho)}\right]^\rho\Pr\left(\boldsymbol{r}_{(\ell n-1)}\middle|\,\boldsymbol{x}_{(\ell n-1)}\right)^{1/(1+\rho)}\Pr\left(\boldsymbol{r}_{(\ell n,\beta n-1)}\middle|\,\boldsymbol{x}_{(\ell n,\beta n-1)}\right)}\\
&= \sum_{\boldsymbol{r}_{(\beta n-1)}\in\Re^{\beta n-1}}\overline{\left[\sum_{\tilde{\boldsymbol{x}}_{(\beta n-1)}\in\mathscr{C}_\beta}\Pr\left(\boldsymbol{r}_{(\beta n-1)}\middle|\,\tilde{\boldsymbol{x}}_{(\beta n-1)}\right)^{1/(1+\rho)}\right]^\rho\Pr\left(\boldsymbol{r}_{(\ell n-1)}\middle|\,\boldsymbol{x}_{(\ell n-1)}\right)^{1/(1+\rho)}\Pr\left(\boldsymbol{r}_{(\ell n,\beta n-1)}\middle|\,\boldsymbol{x}_{(\ell n,\beta n-1)}\right)}\\
&\overset{(a)}{=} \sum_{\boldsymbol{r}_{(\beta n-1)}\in\Re^{\beta n-1}}\overline{\left[\sum_{\tilde{\boldsymbol{x}}_{(\beta n-1)}\in\mathscr{C}_\beta}\Pr\left(\boldsymbol{r}_{(\beta n-1)}\middle|\,\tilde{\boldsymbol{x}}_{(\beta n-1)}\right)^{1/(1+\rho)}\right]^\rho}\,\overline{\Pr\left(\boldsymbol{r}_{(\ell n-1)}\middle|\,\boldsymbol{x}_{(\ell n-1)}\right)^{1/(1+\rho)}\Pr\left(\boldsymbol{r}_{(\ell n,\beta n-1)}\middle|\,\boldsymbol{x}_{(\ell n,\beta n-1)}\right)}\\
&\overset{(b)}{\leq} \sum_{\boldsymbol{r}_{(\beta n-1)}\in\Re^{\beta n-1}}\left[\sum_{\tilde{\boldsymbol{x}}_{(\beta n-1)}\in\mathscr{C}_\beta}\overline{\Pr\left(\boldsymbol{r}_{(\beta n-1)}\middle|\,\tilde{\boldsymbol{x}}_{(\beta n-1)}\right)^{1/(1+\rho)}}\right]^\rho\overline{\Pr\left(\boldsymbol{r}_{(\ell n-1)}\middle|\,\boldsymbol{x}_{(\ell n-1)}\right)^{1/(1+\rho)}\Pr\left(\boldsymbol{r}_{(\ell n,\beta n-1)}\middle|\,\boldsymbol{x}_{(\ell n,\beta n-1)}\right)},\\
&= (|\mathscr{C}_\beta|)^\rho\sum_{\boldsymbol{r}_{(\beta n-1)}\in\Re^{\beta n-1}}\left[\overline{\Pr\left(\boldsymbol{r}_{(\beta n-1)}\middle|\,\tilde{\boldsymbol{x}}_{(\beta n-1)}\right)^{1/(1+\rho)}}\right]^\rho\overline{\Pr\left(\boldsymbol{r}_{(\ell n-1)}\middle|\,\boldsymbol{x}_{(\ell n-1)}\right)^{1/(1+\rho)}\Pr\left(\boldsymbol{r}_{(\ell n,\beta n-1)}\middle|\,\boldsymbol{x}_{(\ell n,\beta n-1)}\right)},
\end{aligned}$$

where $(a)$ holds since labels $\boldsymbol{x}_{(\ell n-1)}$ and any labels in $\mathscr{C}_\beta$ are selected independently, and $(b)$ is valid due to Jensen's inequality with $\rho\leq 1$. Finally, by using the terminologies in Section II-B, and noting that the number of random labels is at most $|\mathscr{C}_\beta|\leq I^{k\beta}=e^{n\beta R}$, we obtain:

$$\begin{aligned}
\overline{\xi(\ell,\beta)} &\leq (|\mathscr{C}_\beta|)^\rho\left[\sum_{j=1}^J\left(\sum_{i=1}^I p_iP_{ji}^{1/(1+\rho)}\right)^{1+\rho}\right]^{\ell n}\left[\sum_{j=1}^J\left(\sum_{i=1}^I p_iP_{ji}\right)\left(\sum_{i=1}^I p_iP_{ji}^{1/(1+\rho)}\right)^\rho\right]^{(\beta-\ell)n}\\
&\leq e^{n\rho R\beta}\left[\sum_{j=1}^J\left(\sum_{i=1}^I p_iP_{ji}^{1/(1+\rho)}\right)^{1+\rho}\right]^{\ell n}\left[\sum_{j=1}^J\left(\sum_{i=1}^I p_iP_{ji}\right)\left(\sum_{i=1}^I p_iP_{ji}^{1/(1+\rho)}\right)^\rho\right]^{(\beta-\ell)n}\\
&= \exp\left\{-\ell n\left[-\rho R+E_0(\rho,\boldsymbol{p})\right]\right\}\exp\left\{-(\beta-\ell)n\left[-\rho R+E_1(\rho,\boldsymbol{p})\right]\right\}, \qquad (11)
\end{aligned}$$

where $E_0(\rho,\boldsymbol{p})$ is defined in (3), and

$$E_1(\rho,\boldsymbol{p})\triangleq-\log\left[\sum_{j=1}^J\left(\sum_{i=1}^I p_iP_{ji}\right)\left(\sum_{i=1}^I p_iP_{ji}^{1/(1+\rho)}\right)^\rho\right].$$

It is obvious that $E_1(\rho,\boldsymbol{p})\geq E_0(\rho,\boldsymbol{p})$ with equality holds if, and only if, either $\rho=0$ or there exists $j\in[1,J]$ such that $P_{ji}=1$ for every $1\leq i\leq I$; thus, $E_1(\rho)\triangleq E_1(\rho,\boldsymbol{p}^*)\geq E_0(\rho)\triangleq\max_{\boldsymbol{p}}E_0(\rho,\boldsymbol{p})$, where $\boldsymbol{p}^*$ is the input distribution that maximizes $E_0(\rho,\boldsymbol{p})$. An example for functions $E_0(\rho)$ and $E_1(\rho)$ over the BSC with crossover probability 0.045 is given in Fig. 4.

Inequality (11) provides an upper probability bound for a top path ending at level $\ell$ being early-eliminated due to $\ell_{\max} = \beta$. Based on (11), we can proceed to derive the bound for the probability $P_{e,E}$ that an incorrect codeword is claimed at the end of the sequential search because the correct path is early-eliminated during the decoding process.

Without loss of generality, assume that the all-zero codeword $\mathbf{0}$ is transmitted. Then,

$$
\begin{aligned}
P_{e,E} = P_{e,E}(\Delta) &\leq \Pr\left(\bigcup_{\ell=1}^{L-\Delta} \mathbf{0}_{n\ell-1} \text{ is early-eliminated}\right) \\
&\leq \sum_{\ell=1}^{L-\Delta} \Pr(\mathbf{0}_{n\ell-1} \text{ is early-eliminated}) \\
&\leq \sum_{\ell=1}^{L-\Delta} \exp\left\{-\ell n\left[-\rho R + E_0(\rho)\right]\right\} \exp\left\{-\Delta n\left[-\rho R + E_1(\rho)\right]\right\},
\end{aligned}
$$

where the last inequality follows from (11) by taking $\boldsymbol{p} = \boldsymbol{p}^*$, and the observation that $\beta - \ell \geq \Delta$. Denoting $\lambda \triangleq E_0(\rho) - \rho R$, we continue the derivation from (12):

$$
\begin{aligned}
P_{e,E} &\leq \exp\left\{-\Delta n\left[-\rho R + E_1(\rho)\right]\right\} \sum_{\ell=1}^{L-\Delta} \exp\left\{-\ell n \lambda\right\} \\
&\leq \exp\left\{-\Delta n\left[-\rho R + E_1(\rho)\right]\right\} \sum_{\ell=1}^{\infty} \exp\left\{-\ell n \lambda\right\} \\
&= K_n \exp\left\{-\Delta n\left[-\rho R + E_1(\rho)\right]\right\}, \qquad (12)
\end{aligned}
$$

where $K_n = e^{-n\lambda}/(1 - e^{-n\lambda})$ is a constant, independent of $\Delta$. Consequently,

$$
\liminf_{n\to\infty} -\frac{1}{n}\log P_{e,E} \geq \Delta[-\rho R + E_1(\rho)] + \lambda \geq \Delta[-\rho R + E_1(\rho)],
$$

subject to $E_0(\rho) > \rho R$ with $0 \leq \rho \leq 1$, which immediately implies:

$$
\liminf_{n\to\infty} -\frac{1}{n}\log P_{e,E} \geq \Delta \cdot E_{el}(R),
$$

where $E_{el}(R) \triangleq \max_{\{\rho\in[0,1]\ :\ E_0(\rho)>\rho R\}}[-\rho R + E_1(\rho)]$. Following similar argument as Forney, we conclude that the additional error due to early elimination in the MLSDA becomes exponentially negligible if

$$
\Delta \cdot E_{el}(R) > (m+1)E_c(R) \quad \text{or equivalently} \quad \Delta/(m+1) > E_c(R)/E_{el}(R) \qquad (13)
$$

for code rates above the channel cutoff rate.

## V. Numerical and Simulation on Early Elimination Windows for Binary Symmetric Channels

For the binary symmetric channel (BSC) with crossover probability $\epsilon$,

$$E_0(\rho) = \rho\log(2) - (1+\rho)\log\left[\epsilon^{1/(1+\rho)} + (1-\epsilon)^{1/(1+\rho)}\right],$$

and

$$E_1(\rho) = \rho\log(2) - \rho\log\left[\epsilon^{1/(1+\rho)} + (1-\epsilon)^{1/(1+\rho)}\right].$$

By choosing $\epsilon = 0.045$ and $\epsilon = 0.095$ to respectively approach the cutoff rates $1/2$ and $1/3$, it can be derived from (13) that the suggested early elimination windows are:

$$\Delta > \frac{0.500}{0.250} \times (m+1) \approx 2.00(m+1) \text{ for rate 1/2 codes} \tag{14}$$

and

$$\Delta > \frac{0.334}{0.333} \times (m+1) \approx 1.00(m+1) \text{ for rate 1/3 codes.} \tag{15}$$

The exponent functions $E_{el}(R)$ and $E_c(R)$ for the above BSCs are plotted in Figs. 5 and 6, respectively. Conditions (14) and (15) indicate that for (2,1,6) and (3,1,8) convolutional codes, respectively taking $\Delta = 15$ and $\Delta = 10$ should suffice to result in negligible performance degradation at the cutoff rates. Simulations are then performed and summarized in Figs. 7 and 8 to examine the analytical results.

It can be observed from Fig. 7 that the MLSDA with early elimination window $\Delta = 15$ does exhibit negligible performance degradation for all $E_b/N_0$'s simulated, where we take $\epsilon = \frac{1}{2}\text{erfc}(\sqrt{E_b/N_0})$ as a convention, and erfc$(\cdot)$ is the complementary error function. This result exactly coincides with our theoretical analysis in (14).

From Fig. 8, we observe that it requires an early elimination window $\Delta = 12$ in order to maintain the maximum-likelihood performance for all signal-to-noise ratios simulated, and the analytical $\Delta = 10$ can only provide negligible performance degradation for $E_b/N_0 \leq 3$ dB. For example, the bit error rates (BERs) for $\Delta = 10$ and $\Delta = 20$ are $1.40 \times 10^{-3}$ and $1.04 \times 10^{-3}$, respectively, at $E_b/N_0 = 4$ dB. This hints that when channels become less noisy, the contribution of the errors due to early elimination is a little more dominant to the overall performance than that of the maximum-likelihood decision errors. Hence, a slightly larger early elimination window than the analytical one $\Delta = 10$ is necessary to bring down the dominance of the early elimination errors.

The above "under-estimation" of the early elimination window is quite different from Forney's estimation of the sufficient path truncation window, for which an over-estimation is usually resulted. The under-estimation of early elimination windows, as well as the over-estimation of path truncation windows, can be reasoned from the multiplicative constants in the error probability bounds. In Forney's argument, it requires:

$$\exp\left\{-n\tau E_r(R)\right\} \ll \frac{1}{1-2^{-\lambda/R}} \exp\left\{-n(m+1)E_0(R)\right\}. \tag{16}$$

where $q = 2$ for the BSCs. In our derivation, we demand:

$$\frac{e^{-n\lambda}}{1-e^{-n\lambda}} \exp\left\{-n\Delta E_{el}(R)\right\} \ll \frac{1}{1-2^{-\lambda/R}} \exp\left\{-n(m+1)E_0(\rho)\right\},$$

or equivalently,

$$\exp\left\{-n\Delta E_e(R)\right\} \ll 2^{\lambda/R} \exp\left\{-n(m+1)E_0(\rho)\right\}, \tag{17}$$

since $R = (1/n)\log(2)$ nat/symbol for $(n, 1, m)$ binary convolutional codes. By taking the optimal $\rho$ that yields the best exponent, i.e., $\lambda = E_0(\rho) - \rho R = 0$, the multiplicative constant $1/(1-2^{-\lambda/R})$ in (16) approaches infinity,[4] while the multiplicative constant $2^{\lambda/R}$ in (17) reduces to $1$. This explains the reason why our estimate of early elimination windows tends to be exact (e.g., for the simulated (2,1,6) convolutional code) or under-estimated (e.g., for the simulated (3,1,8) convolutional code), while Forney's estimate of path truncation windows is often larger than necessary.

[4] Zigangirov [20] has provided a tighter decoding error bound for $P_{e,c}$ by simultaneously minimizing the multiplicative constant, $(q-1)/(1-q^{-\lambda/R})$, and the exponential term, $\exp\{-n(m+1)E_0(\rho)\}$. Zigangirov's probability bound, however, gives the same asymptotic error exponent as (5) even it is better for finite $n(m+1)$.

The simulated reduction of decoding complexity due to early elimination is plotted in Fig. 9. The decoding complexity is measured by the average number of branch metric computations per information bit. From Fig. 9, we found that the computational complexity is significantly reduced at medium signal-to-noise ratios. For example, at $E_b/N_0 = 6$ dB, the MLSDA with early elimination window $\Delta = 15$ only requires 5.73 metric computations per information bit. This number is around one eleventh of 62.7, which is the average computational complexity of the MLSDA without early elimination.

Figure 10 depicts the the “$99.9\%$ Open Stack size” that is defined as the Open Stack size to complete $99.9\%$ of the simulation runs without stack overflow. The $99.9\%$ Open Stack size is equivalent to the required Open Stack size such that the stack overflow probability is less than $10^{-3}$. It should be mentioned beforehand that the Closed Stack can be implemented by simply adopting a bitmap table that uses one bit per node to identify whether a node has once been the end node of the top path of the Open Stack. As a result, the Closed Stack consumes much less memory than the Open Stack, and hence, only the size of the Open Stack is necessarily simulated.

Several observations can be made from Fig. 10. Firstly, the Open Stack size for the MLSDA without early elimination may grow large at medium SNRs. As an example, the MLSDA without early elimination requires an Open Stack size of 2877 in order to satisfy $99.9\%$ of the $10^5$ simulation runs at $E_b/N_0 = 6$ dB. Secondly, unlike the conventional sequential decoding that operates on a code tree, the memory requirement of the Open Stack for the trellis-based MLSDA without early elimination also decreases at low SNRs. This is because although the decoding process visits almost all trellis nodes at low SNRs, most of the visited paths will be directly eliminated by path merging (cf. Step 3 of the trellis-based MLSDA). Thirdly, early elimination modification flattens the line below which the stack size is sufficient to fulfil $99.9\%$ of the simulation runs, and the $99.9\%$ Open Stack size becomes less relevant to SNRs. Meanwhile, only one ninth of the $99.9\%$ Open Stack size, specifically 318, is required at $E_b/N_0 = 6$ dB when early elimination is applied to the MLSDA.

A final observation on Fig. 10 is that the 99.9% Open Stack size of the MLSDA without early elimination converges to its smallest possible value $(L+1) = 201$ when $E_b/N_0$ is beyond 11.5 dB. It can therefore be expected that the sequential decoding search of the MLSDA shall go all the way to the terminal node of the code trellis at $E_b/N_0 = 11.5$ dB. The early elimination modification however reduces the SNR at which the 99.9% Open Stack size converges downto 9.5 dB by eliminating all the pathes deviating from the path corresponding to the final decoding decision.

Finally, we should emphasize that the 99.9% Open stack size is different from the minimum stack size required to introduce negligible performance degradation. The traditional way to handle stack overflow under finite stack size is to eliminate the paths at the bottom of the stack in order to make rooms for the newly arrived paths [13]. For example, by denoting OPENMAX as the boundary of the stack size, beyond which the newly added path will push out the path with the largest metric when the stack is full [8], Fig. 11 shows that OPENMAX= 2560 is sufficient to maintain near-ML performance for the MLSDA without early elimination. This number is smaller than 2877, i.e., the 99.9% Open Stack size at $E_b/N_0 = 6$ dB. The introduction of the early elimination can again lower OPENMAX downto 256.

## VI. Concluding Remarks and Future Work

In this work, we propose to improve the computational complexity and memory requirement of the maximum-likelihood sequential-search decoding algorithm by early elimination. The random coding analysis of the sufficient early elimination window for negligible performance degradation, as well as the subsequent simulations, confirms our anticipated improvement. Since the MLSDA with early elimination is justified to suit applications that dictate a near-ML software decoder with limited support in computational power and memory, a future work of practical interest will be to apply the MLSDA with early elimination to the “supper-code” for joint multi-path channel equalization and convolution decoding [10].

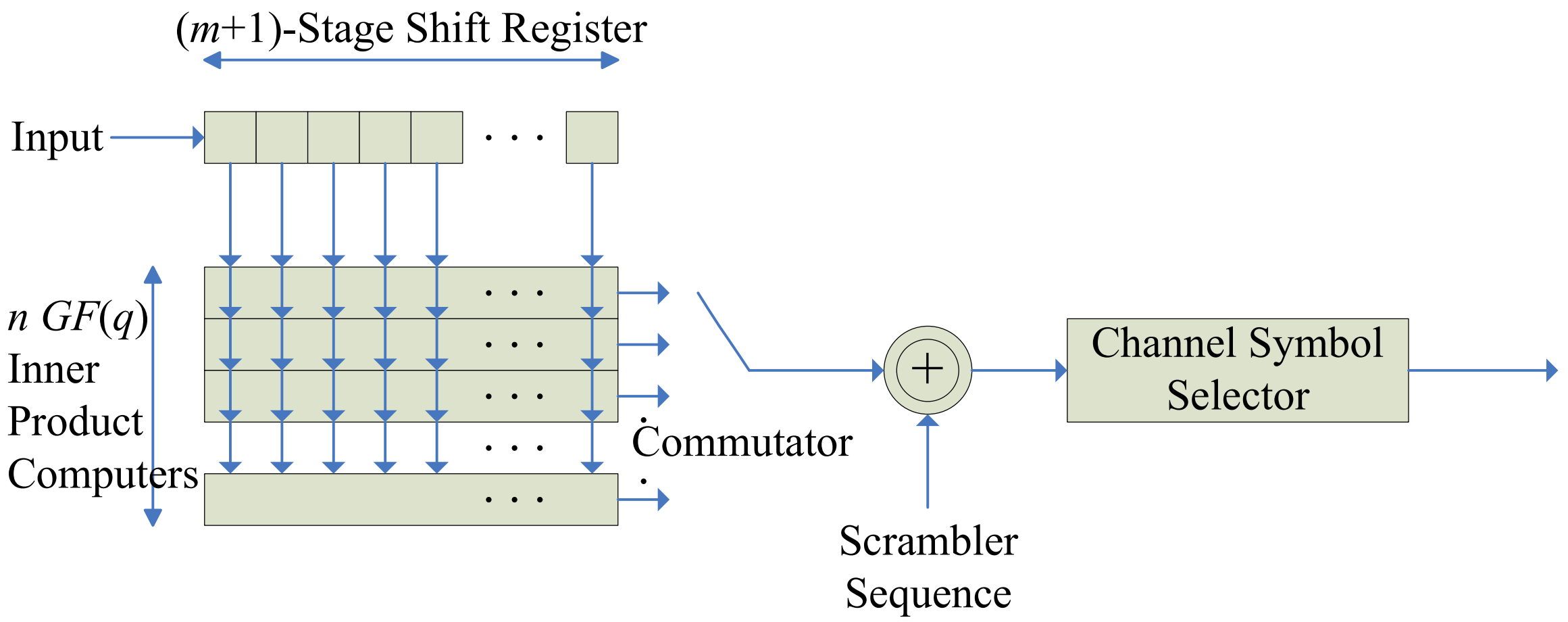

Fig. 1. Single-input $n$-output encoder model considered in [17]. All elements are in $GF(q)$, where $q$ is either a prime or a power of a prime.

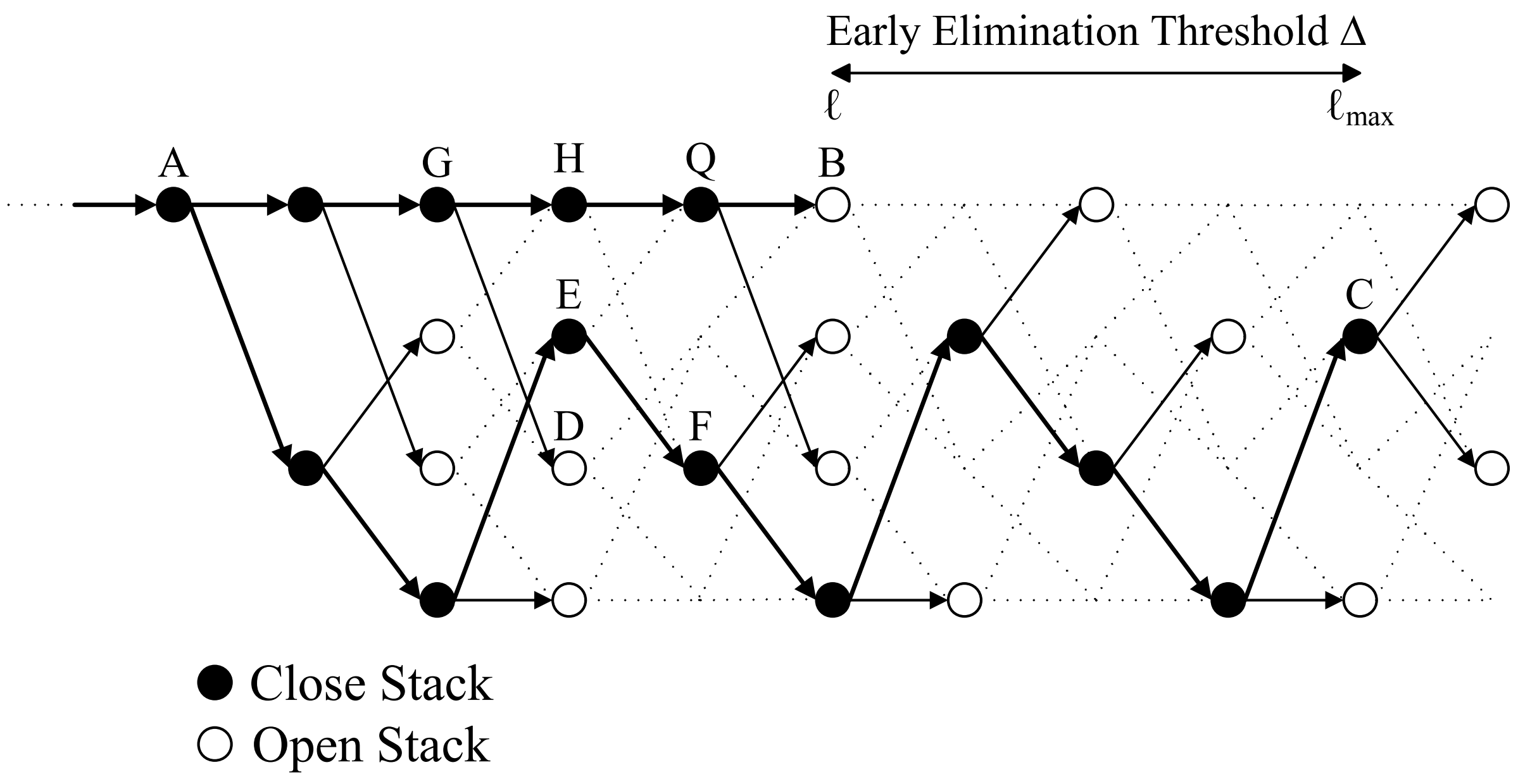

Fig. 2. Early elimination window $\Delta$ in the trellis-based MLSDA.

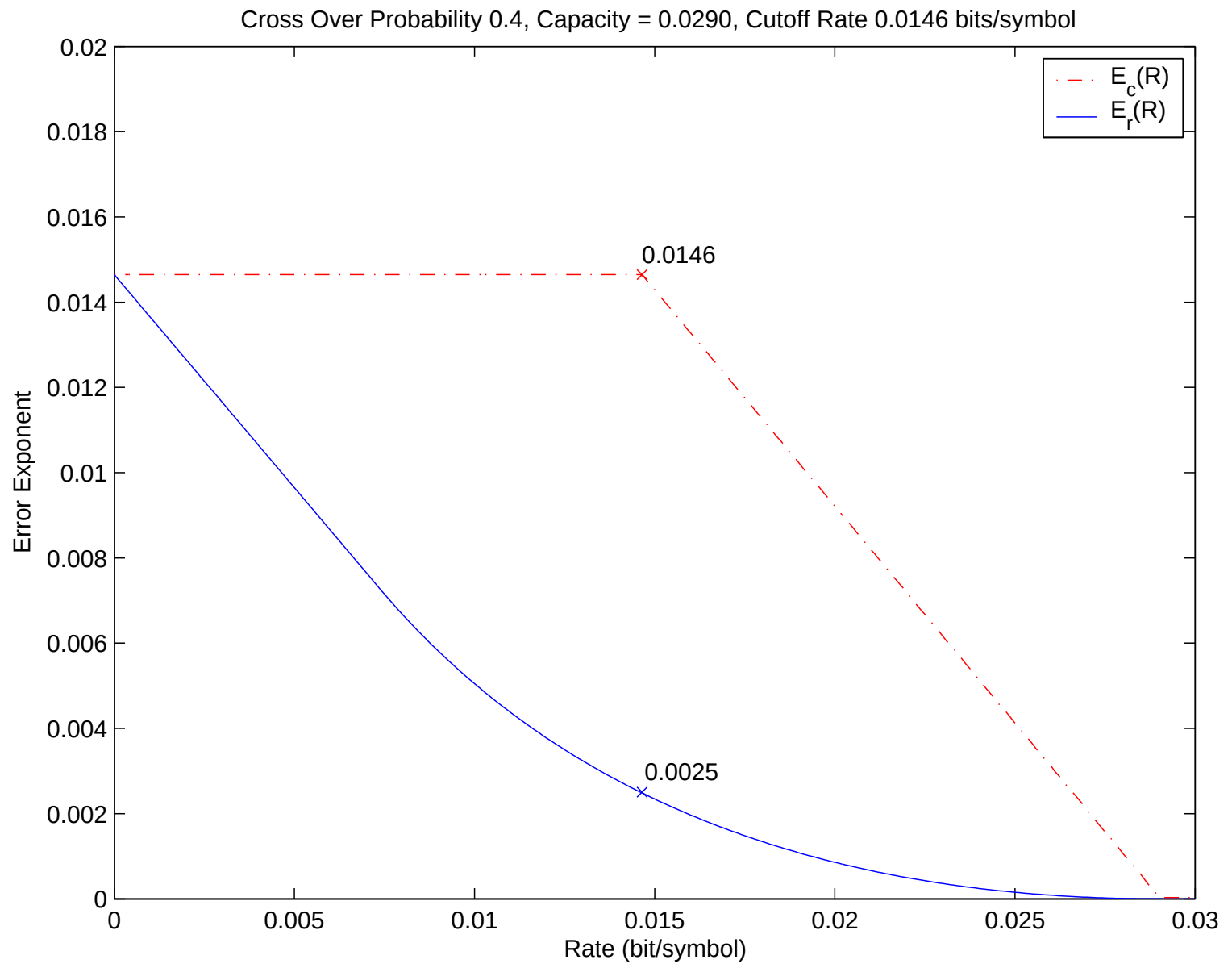

Fig. 3. Exponent lower bound $E_r(R)$ of the additional error due to path truncation and exponent $E_c(R)$ of the maximum-likelihood decoding error for time-varying convolutional codes (without path truncation) under the BSC with crossover probability $0.4$.

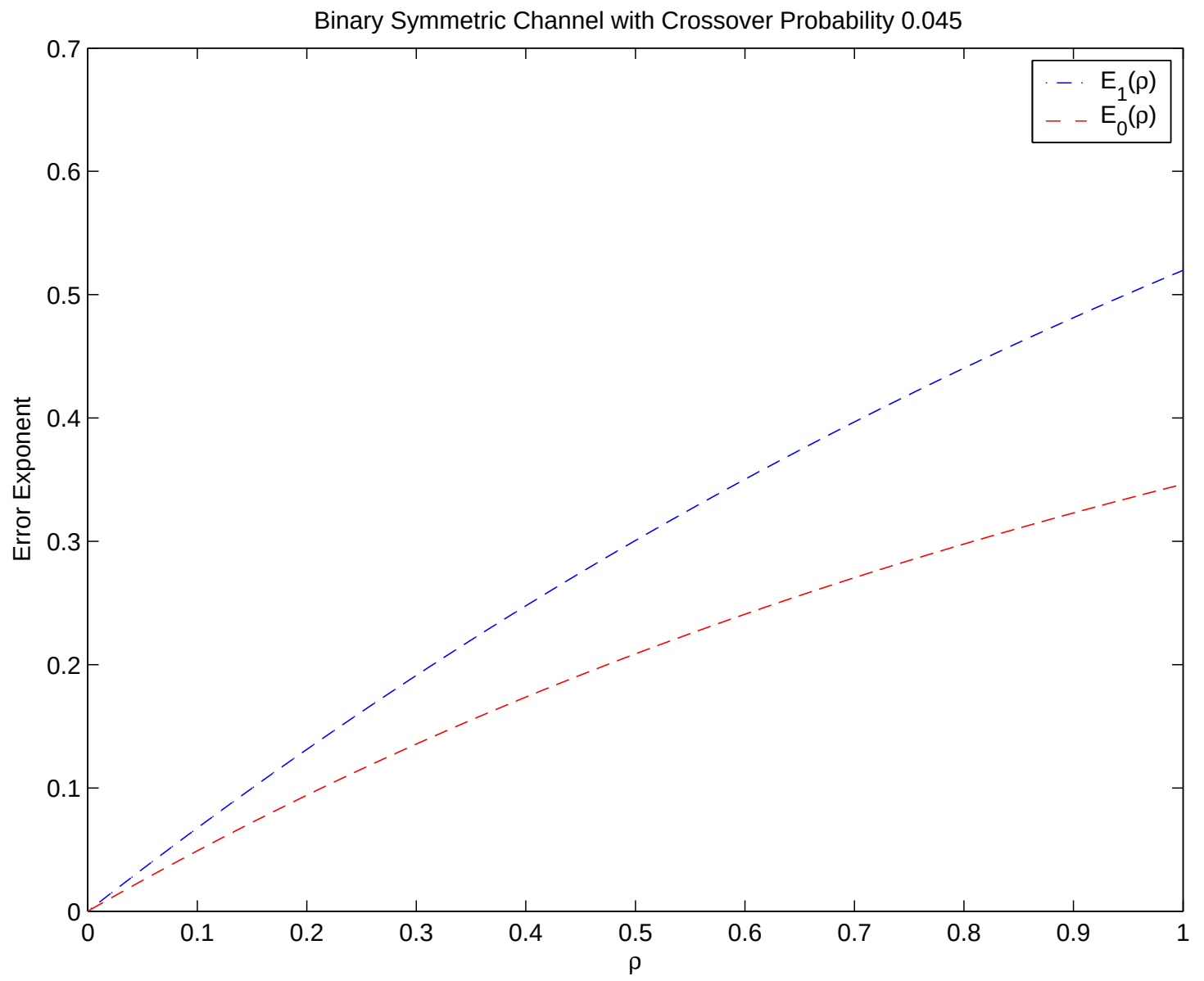

Fig. 4. Functions $E_0(\rho)$ and $E_1(\rho)$ under the BSC with crossover probability $0.045$.

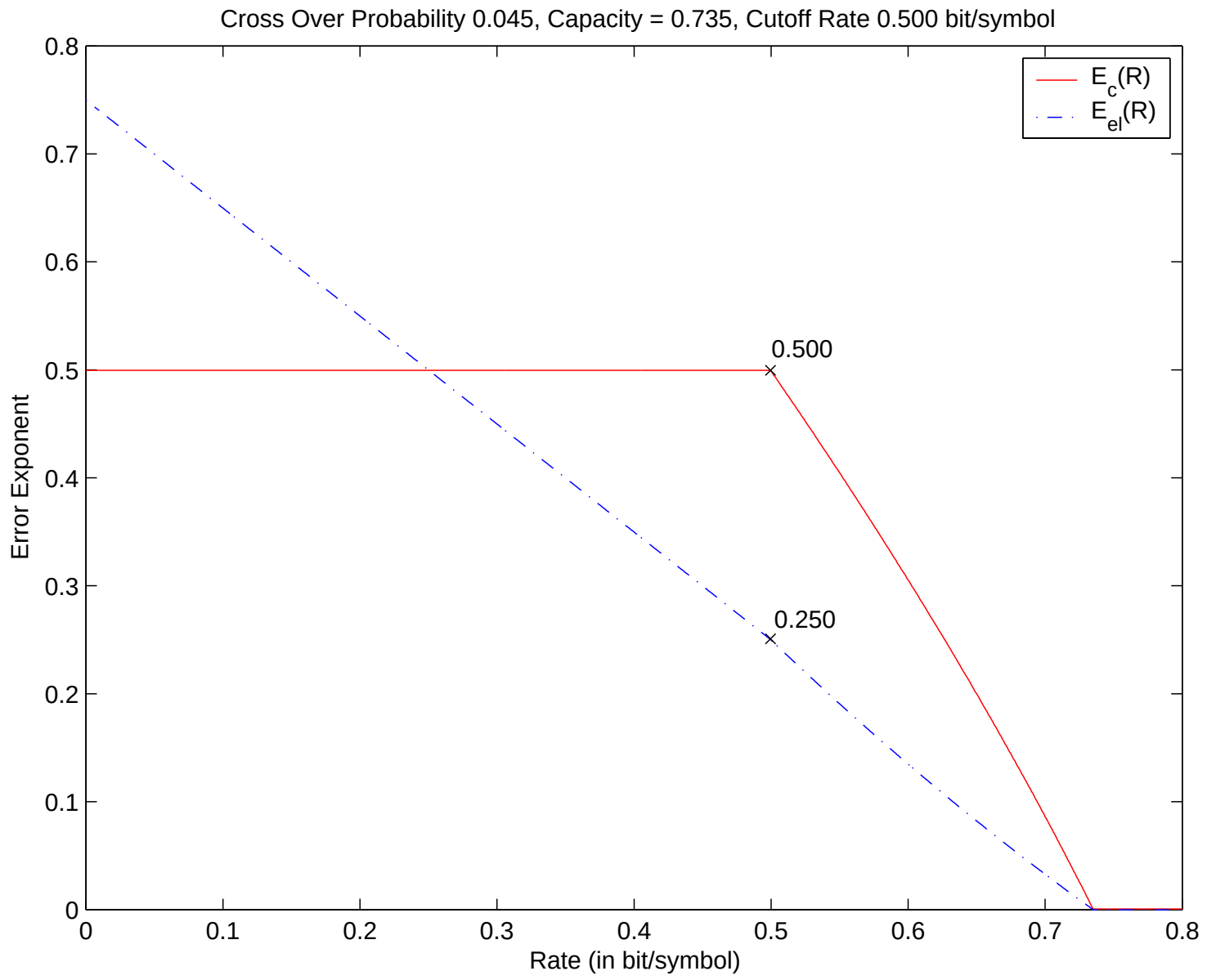

Fig. 5. Exponent lower bound $E_{el}(R)$ of the additional error due to early elimination and exponent $E_c(R)$ of the maximum-likelihood decoding error for time-varying convolutional codes (without early elimination) under the BSC with crossover probability $0.045$.

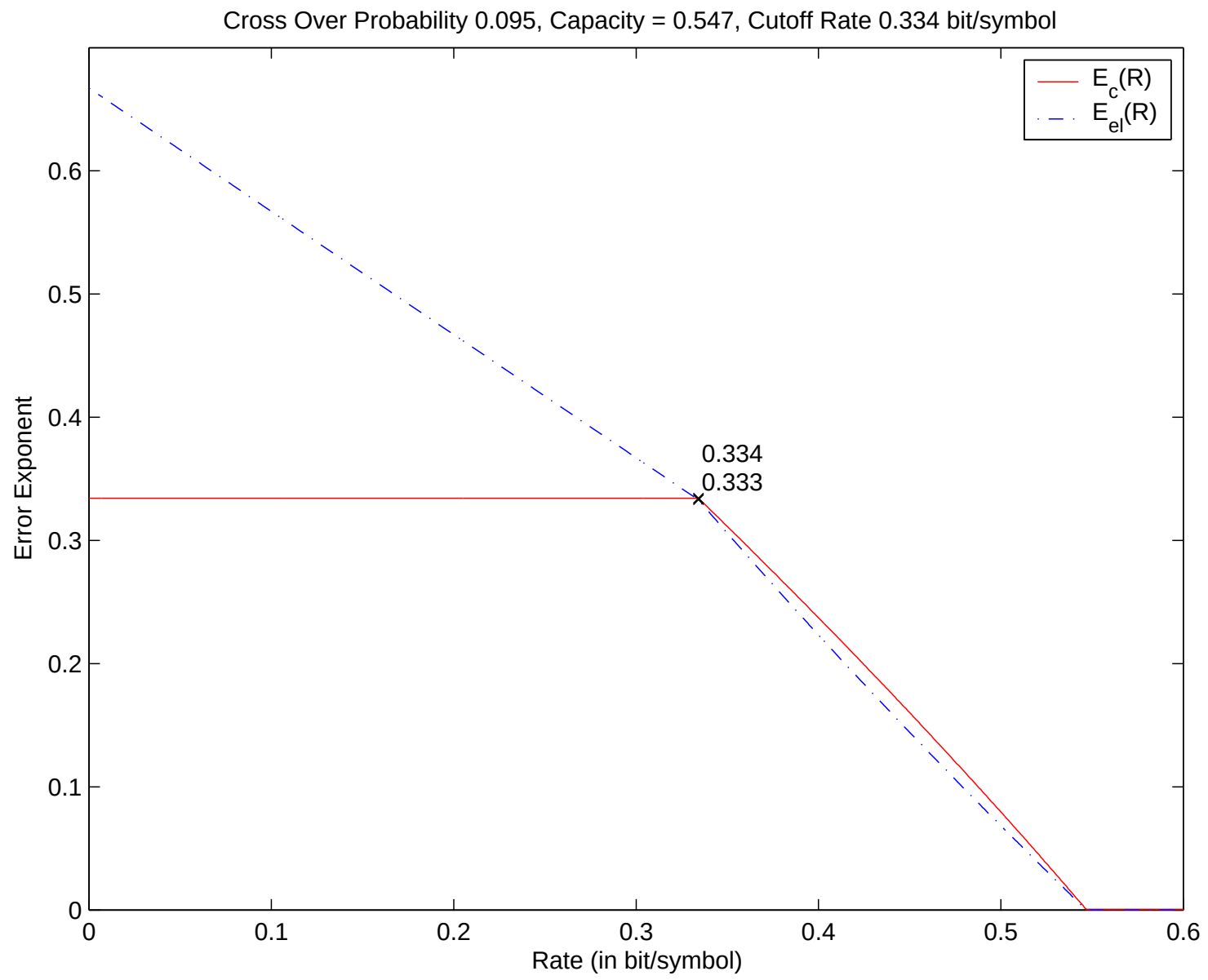

Fig. 6. Exponent lower bound $E_{el}(R)$ of the additional error due to early elimination and exponent $E_c(R)$ of the maximum-likelihood decoding error for time-varying convolutional codes (without early elimination) under the BSC with crossover probability $0.095$.

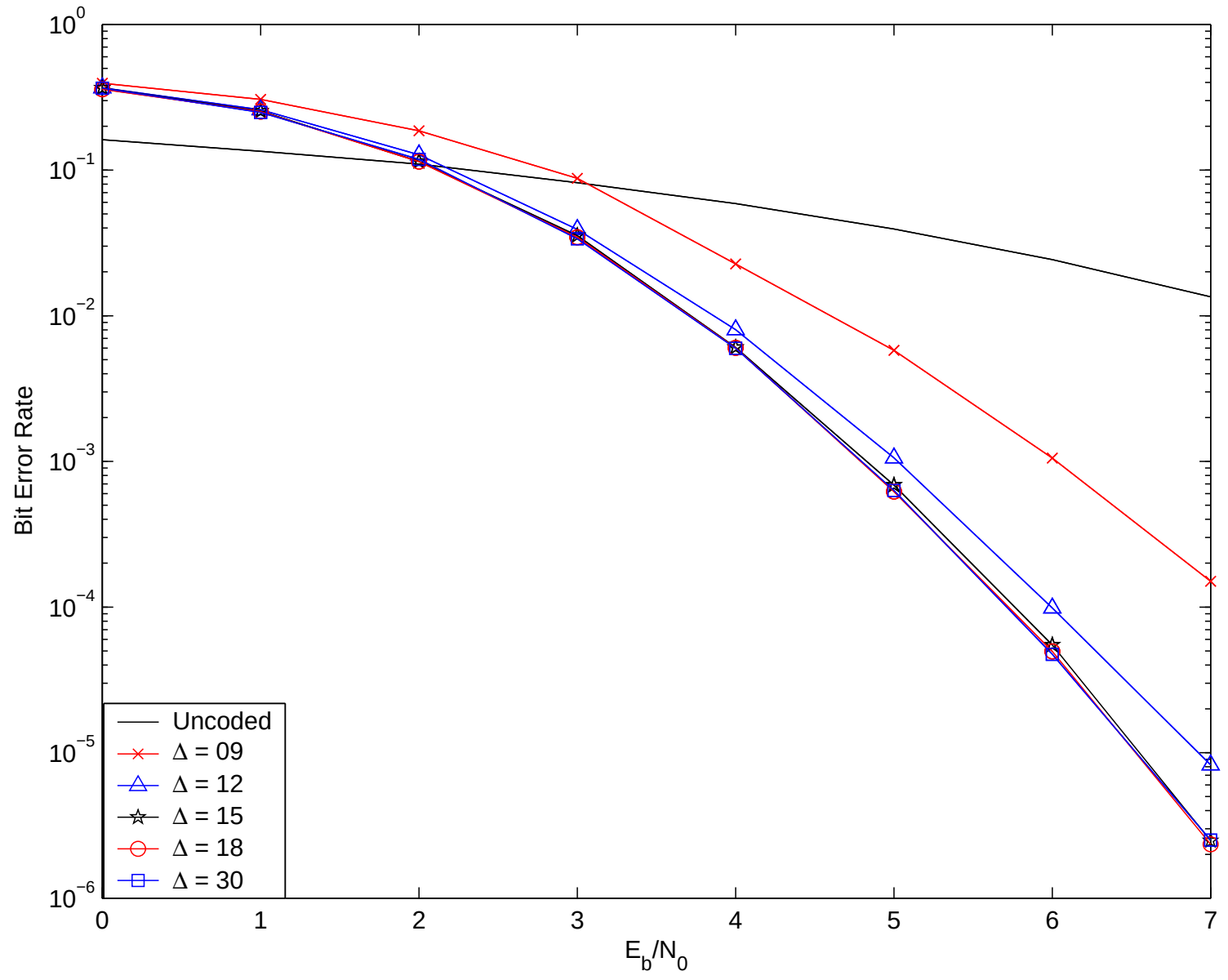

Fig. 7. Performance for (2,1,6) convolutional codes under different early elimination windows. The generator polynomial for (2,1,6) convolutional codes is [554 774] in octal. The message length is infinite, and the backsearch limit is set to be $\tau = 40$.

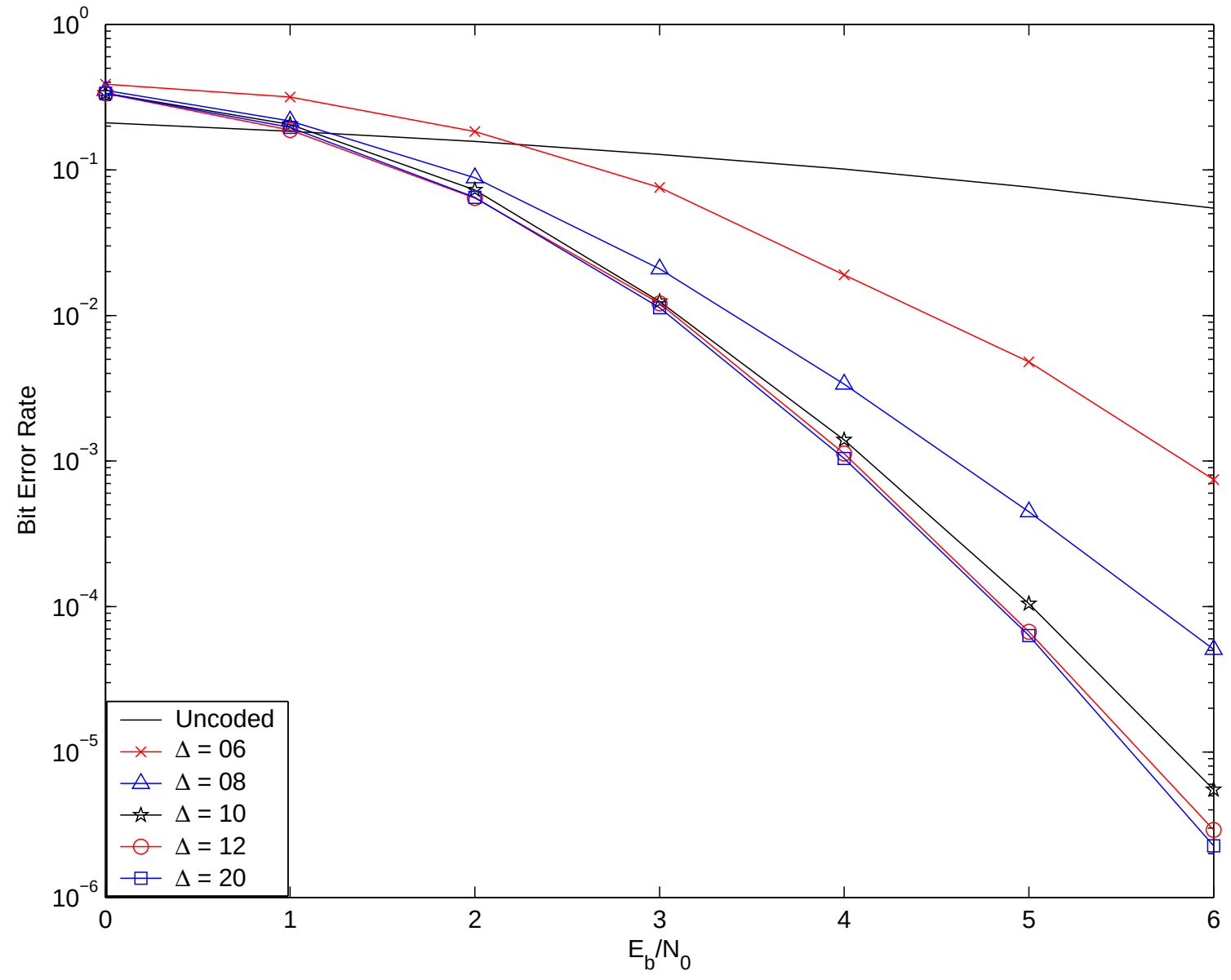

Fig. 8. Performance for (3,1,8) convolutional codes under different early elimination windows. The generator polynomial for (3,1,8) convolutional codes is [557 663 711] in octal. The message length is infinite, and the backsearch limit is set to be $\tau = 52$.

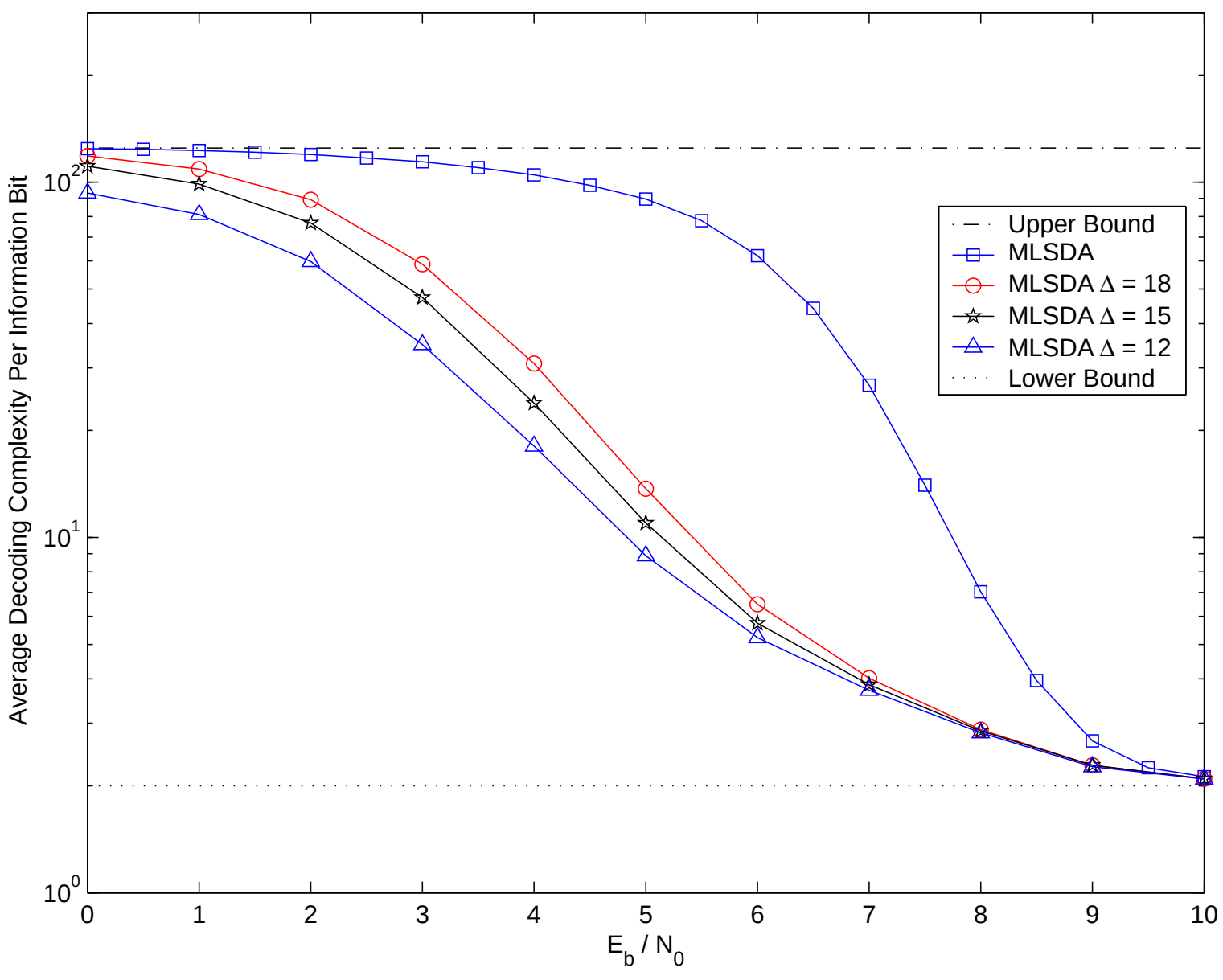

Fig. 9. Average branch metric computations per information bit for (2,1,6) convolutional codes under different early elimination windows. The message length is $L = 200$, and no backsearch limit is utilized. Notably, the decoding complexity is upper-bounded by $2[2^m L - (m-2)2^m - 2]/L \approx 125.4$, and is lower-bounded by $(2L+m)/L \approx 2$.

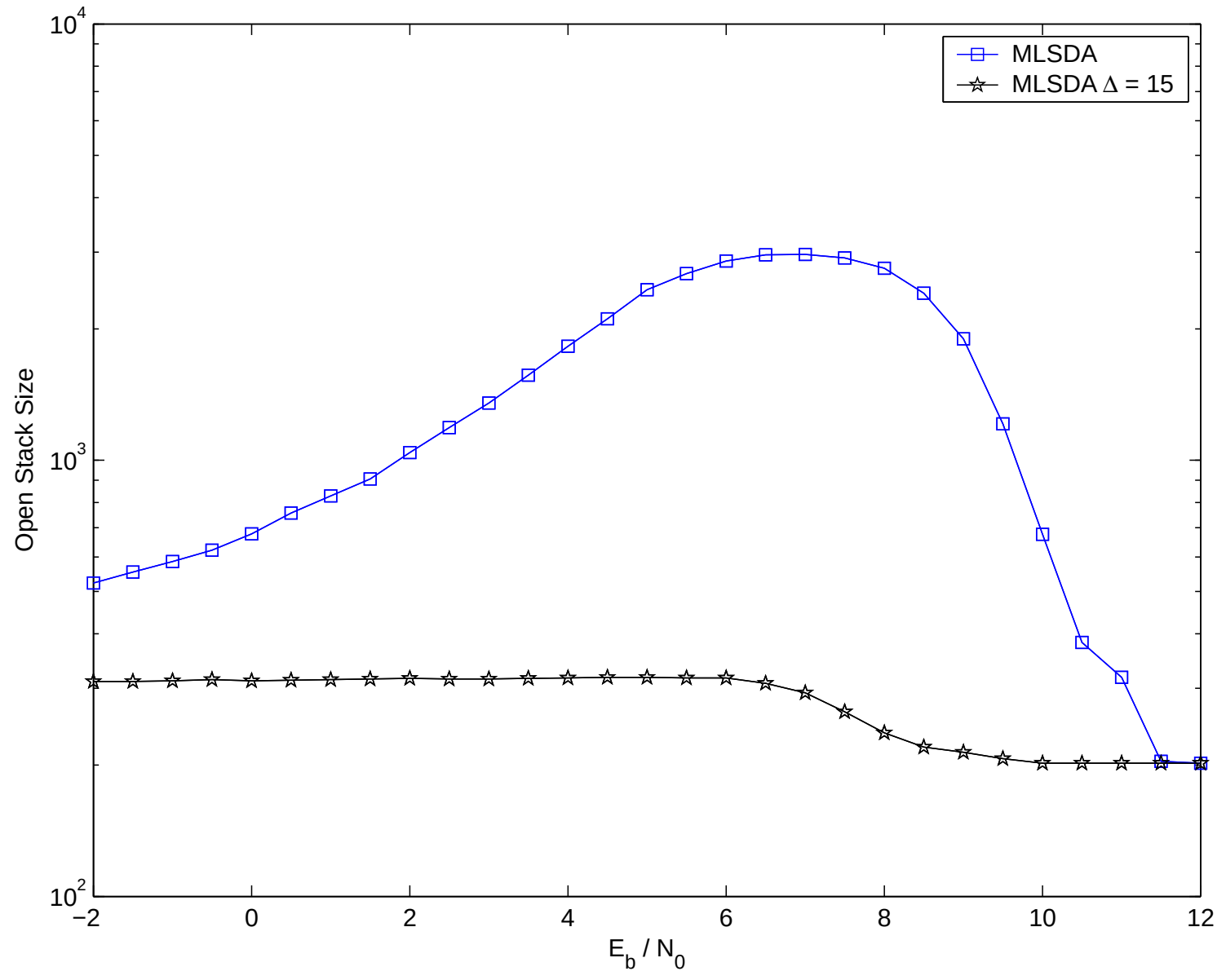

Fig. 10. The 99.9% Open Stack size of the MLSDA with and without early elimination for (2,1,6) convolutional codes. The message length $L$ is $200$, and no back search limit is utilized.

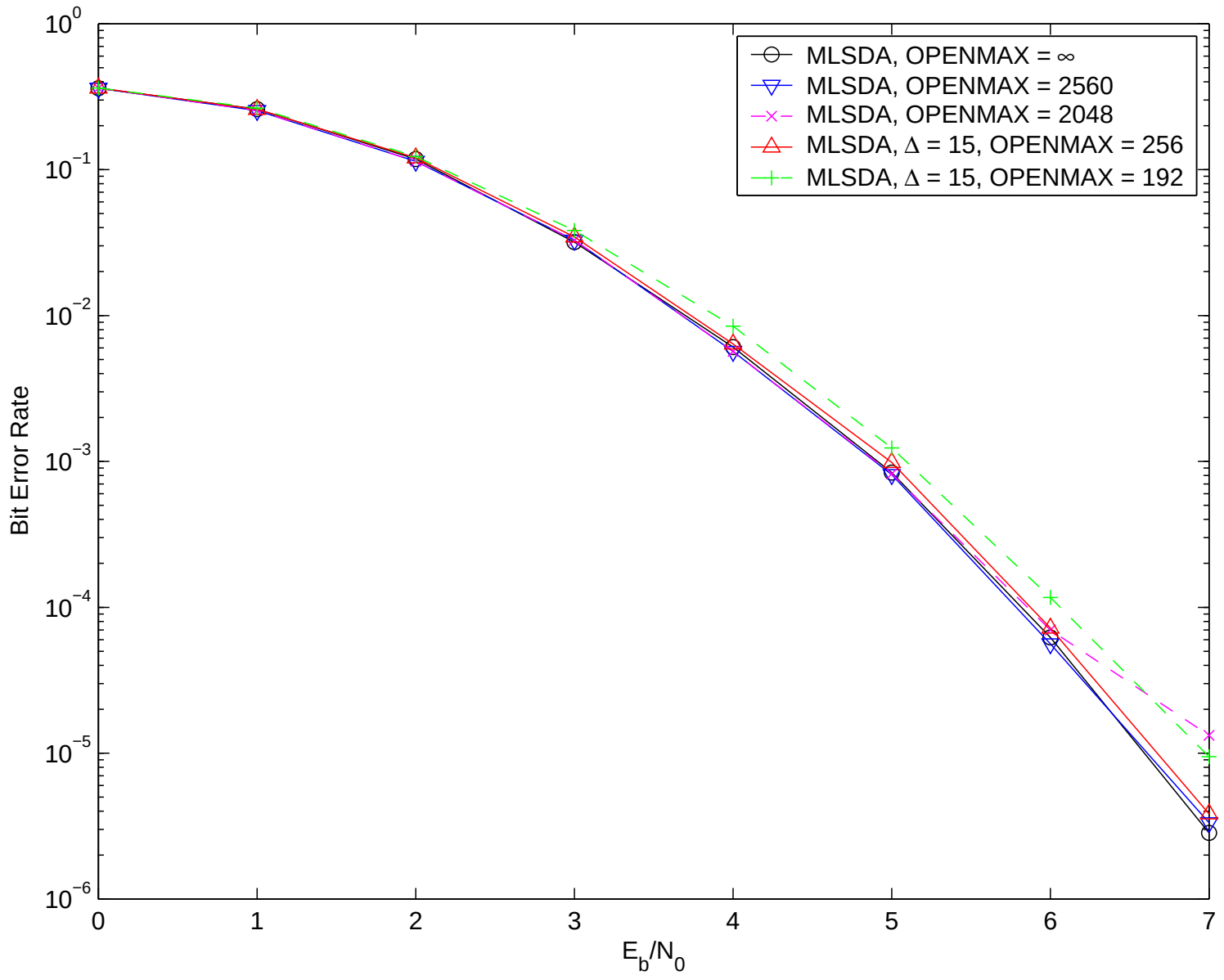

Fig. 11. Performances of the MLSDA with and without early elimination subject to the constraint of finite stack size for (2,1,6) convolutional codes. OPENMAX is the upper boundary of the Open Stack size. The message length is $L = 200$, and no back search limit is utilized.